**Silk reinforced with graphene or carbon nanotubes spun by spiders**

Emiliano Lepore[1], Francesco Bonaccorso[2,3], Matteo Bruna[2], Federico Bosia[4], Simone Taioli[5,6], Giovanni Garberoglio[5], Andrea. C. Ferrari[2], Nicola Maria Pugno[1,7,8*]

[1] *Laboratory of Bio-inspired & Graphene Nanomechanics, Department of Civil, Environmental and Mechanical Engineering, University of Trento, Via Mesiano 77, 38123 Trento, Italy.*
[2] *Cambridge Graphene Centre, University of Cambridge, 9 JJ Thomson Avenue, CB3 0FA, Cambridge UK.*
[3] *Istituto Italiano di Tecnologia, Graphene Labs, Via Morego 30, 16163 Genova, Italy.*
[4] *Department of Physics and "Nanostructured Interfaces and Surfaces" Centre, Università di Torino, Via P. Giuria 1, 10125, Torino, Italy.*
[5] *European Centre for Theoretical Studies in Nuclear Physics and Related Areas (ECT\*), Bruno Kessler Foundation & Trento Institute for Fundamental Physics and Applications (INFN-TIFPA), Trento, Italy.*
[6] *Faculty of Mathematics and Physics, Charles University, Prague, Czech Republic.*
[7] *Centre of Materials and Microsystems, Bruno Kessler Foundation, Via Santa Croce 77, 38122 Trento, Italy.*
[8] *School of Engineering and Materials Science, Queen Mary University, Mile End Road, London E1 4NS, UK.*
[*]*Corresponding author: nicola.pugno@unitn.it*

**The protein matrix and hard tissues of insects[1,2,3,4], worms[5,6,7], ants[8] and spiders[9,10] naturally incorporates metals, such as zinc[1,2,3,5,8,9,10], manganese[2,3,9] and copper[6,7]. This leads to mechanical hardening of teeth[9], jaws[5,6,7], mandibles[1,2,3,4,8,], ovipositors[8] and to an enhancement of silk toughness[10]. Thus, the artificial incorporation of metals, or even insulating or semiconducting materials, into these protein structures could be exploited to obtain a reinforced matrix. A number of groups reported the introduction of metals, such as zinc[1,2,3,5,8,9], titanium[10], aluminium[5], copper[6,7] and lead[6] in the protein structure of spider silk through multiple pulsed vapour-phase infiltrations[10]. This allowed to increase its toughness modulus from 131 MPa[8] up to 1.5 GPa[10]. Biomaterials with increased mechanical or conductive properties could find innovative applications in garment textiles[11] and medical nerve regeneration[12]. It was suggested to coat spider silks with amine-functionalized multi-wall carbon nanotubes, to produce electrically conducting fibres[9], or with cadmium telluride[13], magnetite[1] or gold[1,2] nanoparticles, for fluorescent[13], magnetic[13,14] and electronic applications[13,15]. However, to the best of our knowledge, the incorporation of materials in the inner protein structure of spider silk has not been achieved to date. Here, we report the production of silk incorporating graphene and carbon nanotubes directly by spider spinning, after spraying spiders with the corresponding aqueous dispersions. We observe a significant increment of the mechanical properties with respect to the pristine silk, in terms of fracture strength, Young's and toughness moduli. We measure a fracture strength up to~5.4 GPa, a Young's modulus up to~47.8 GPa and a toughness modulus up to ~2.1 GPa, or 1567 J/g, which, to the best of our knowledge, is the highest reported to date, even when compared to the current toughest knotted fibres[8]. This approach could be extended to other animals and plants and could lead to a new class of bionic materials for ultimate applications.**

Silkworm silks have been widely used by mankind for millennia. Although over 40,000 spider species have been classified to date[16,17,18,19,20], spider silks have been extensively studied only during the last decades[21,22], with an increasing number of publications and patents focusing on their outstanding mechanical[23,24], conducting[2,14], electrical[13,15], biocompatibility[12,25,26] and low immunogenic[19] properties[6]. These make silk potentially important for many applications, from garment textiles[11,19,20] to electrical[21], sensor[13,14,15] and actuating[15,23] devices or medical applications[19,27,28] such as suture threads[29], biomimetic muscles[16], nerve regeneration[12], ligament tissue repair[19], scaffolds[7,20], up to long term applications such as flak jackets (currently limited by silk's large deformability)[3,4,6,9].

Silk is one of the keys to spider's evolutionary success[10] and has been perfected over 400million years of evolution[17,30]. Spider silk is generally described as a semi-crystalline biocompatible composite biopolymer[12,19] and comprises the amino acids alanine, glycine and serine, organized into semi-amorphous helical-elastic $\alpha$-chains and $\beta$-pleated nanocrystals[6,7,9,19,21]. Spider silk is considered the best amongst spun polymer fibres[5,8] in terms of tensile strength and strain, therefore toughness modulus, even when compared with the best performing synthetic fibres, such as ultra-high molecular weight polyethylene or poly(p-phenylene-2,6-benzobisoxazole)[5,19], or with moth silk, which originates from silkworms[7]. The mechanism of spider silk spinning involves a number of biological[25,26], chemical[24,28,31] and physical[24,] processes[17], leading to silk fibres with a Young's modulus ~10GPa[32] and fracture strength of approximately 1 GPa for dragline silk[5].

Unlike the case of the largely available silkworm silks, large-scale spider farming and synthetic production of spider silk still remain to be achieved, due to its complex structure[6] and the territorial and cannibalistic nature of spiders[19,30]. Moreover, naturally spun fibres[24], obtained by forcible spinning[20,33,34], harvesting[19] or extracting spidroin from glands[35], have reduced mechanical characteristics with respect to naturally-spun ones[34] (e.g. the Young's modulus is 11.1 GPa[33] (10.4 GPa[34]) instead of 6.9 GPa[33] (9.2 GPa[34]), the ultimate strain is 0.2%[33] (0.33%[34]) instead of 0.3%[33] (0.47%[34]) and the stress is 1.2 GPa[33] instead of 0.6 GPa[34], due to the $CO_2$ anaesthesia of spiders[33] and the consequent loss of active control of their silk spinning[34]. Research to improve the mechanical[6,23], conductive[2,14] or magnetic[14] properties of spider silk has been limited[14,15,21]. This is due to the difficulty of developing an adequate spinning methodology, balancing extrusion, drawing, yield and purity[19]. From a technological point of view[19], wet-spinning[36], electro-spinning[37,38], hand-drawing[39,40] or microfluidics approaches[15] have been investigated to produce an artificial silk biomaterial at lab scale[36,37,38,39,40], mechanically, structurally or chemically modified with respect to natural one[19,20]. Attempts to improve or modify the mechanical, magnetic and electrical properties of spider silk have been reported, using techniques such as melt-spinning[11], templating[14], powder coating[1,6,9,25], atomic layer deposition[5] and iodine doping[2], but they remain to be adequately perfected, especially at large scale, using naturally spun spider silk fibres.

Here, we study silk directly spun by spiders produced after the exposure of spiders to dispersions of carbon nanotube (CNTS) or graphene (GS). We find that the resulting silk has improved mechanical properties: a fracture strength up to ~5.4 GPa, a Young's modulus up to ~47.8 GPa, and an toughness modulus up to ~2.1 GPa. This is the highest toughness modulus for a fibre, surpassing synthetic polymeric high performance fibres (e.g. Kelvar49$^{TM}$ [30]) and even the current toughest knotted fibers[8].

Two types of SWNTs are used in this study. The first is CoMoCAT SWNTs (named SWNT-1). The second is electric arc discharge SWNT (P2) from Carbon solutions inc. (named SWNT-2). SWNT dispersions are prepared by adding 1mg/10ml weight to volume ratio of each SWNT source to an aqueous solution of 2% w/v sodium deoxycholate (SDC, Sigma-Aldrich) in deionised water (DIW). De-bundling is obtained via ultrasonication using a Branson Ultrasonic Processor for 2 hours (450 kW at 20 kHz). The dispersions are ultracentrifuged using a TH-641 swinging bucket rotor in a Sorvall WX-100 ultracentrifuge at 200,000 x g for 2 hours at 18°C to remove SWNT bundles and other impurities, such as amorphous carbon, catalyst residuals, etc[41].

The supernatant of the two dispersions after the ultracentrifugation step is collected using pipettes and used for the characterization. Graphite flakes are sourced from Sigma Aldrich Ltd. 100mg are dispersed in 10ml water with 2% v/w SDC. The dispersion is then ultrasonicated for 10 hours and subsequently ultracentrifuged, exploiting sedimentation-based separation (SBS) using a TH-641 swinging bucket rotor in a Sorvall WX-100 ultracentrifuge at 5000 rpm for 30 mins. After ultracentrifugation, the supernatant is extracted by pipetting. The concentration of graphitic flakes is determined from the optical absorption coefficient at 660nm, as described in Ref.42. A full optical and spectroscopic characterization of the samples is presented in Supplementary Information (S.I.)

Spiders were selected as described in S.I. Our silk fibres consist of multiple threads of approximately circular cross-section. An average diameter of the silk single thread is determined for each sample, at two different cross-sections along its length. This gives a nearly constant cross-sectional surface area along the fibre length. The number of threads in each fibre is also counted. The cross-sectional surface area of each fibre is calculated by multiplying the mean value of the cross-sectional area of the threads by their total number (*e.g.*, the mean cross-sectional area of threads of sample 1 is 0.202 $\mu m^2$ and the total number of threads is 15, resulting in a cross-sectional surface area of the fibre of sample 1 equal to 3.03 $\mu m^2$).

The presence of nanotubes and graphene is monitored by Raman spectroscopy. Fig. 1 compares the Raman spectra of the reference spider silk (RS) with that collected from spiders exposed to dispersions of graphene (GS) (Fig. 1b) and SWNTs (CNTS) (Fig. 1c,d). Fig. 1a shows an optical image of a suspended fibre. The RS Raman spectrum comprises several peaks in the 1000-1800 $cm^{-1}$ and 2700-3500 $cm^{-1}$ spectral regions. The peaks at ~1088 and 1160$cm^{-1}$ are characteristic of the n(C-C) skeletal band of polypeptide chains[16,26]. Two intense band are also seen at ~1230 $cm^{-1}$, characteristic of amide III groups in π-sheets structured proteins[12,31] and at ~1444$cm^{-1}$ assigned to $CH_2$ bending modes, both bands typically found in the Raman spectrum of spider silk[24]. The peaks at 1615$cm^{-1}$ and 1665$cm^{-1}$ are assigned to n(CO) amide I bands characteristic of the π-sheets configuration for the polypeptide backbone.[12,24]. The Raman peaks in the region 2700-3500 $cm^{-1}$ are typical of C-H and N-H vibrations[31].

Raman spectra are also measured from GS and CNTS samples (see Fig. 2). The intensity of the different spectra is normalized with respect to the C-H band at~2934$cm^{-1}$, the most intense in RS. The normalized RS spectrum is then subtracted from the GS and CNTS ones. Figs. 2a,b show that the spectrum of the graphene flakes in the dispersion is compatible with that of GS. At both 514 and 633nm, the D to G and 2D to G intensity ratios, I(D)/I(G) and I(2D)/I(G), as well as the positions of the G and 2D peaks, Pos(G) and Pos(2D), are very similar. The comparison indicates that graphene detected in GS has a similar level of disorder as the original material. The same holds for the case of CNTS. Figs 2c-f show that the spectra of the original SWNTs and those measured on CNTS are similar, indicating a negligible change in the structural properties.

Nanotensile tests are performed under controlled conditions as described in the S.I. Samples are subjected to traction up to failure at a constant strain rate of 0.1 %/s, consistently with previous studies on spider silk mechanics[1,13,32]. The stress *σ*, ultimate strain *ε*, and Young's modulus *E*, are calculated as $\sigma = F/A_0$, $\varepsilon = \Delta l/l_0$, $E = d\sigma/d\varepsilon|_0$, where *F* is the force measured by our nanotensile system (see S.I. for details), $A_0$ is the cross-sectional surface area of the fibre, and $\Delta l$ is the change in fibre length measured during the test. The area underlying the stress-strain curve corresponds to the energy per unit volume required to break the fibre[23], the so-called toughness modulus, also alternatively given in energy per unit mass, once the energy per unit volume is divided by the density of the material[8]. The variations of the mechanical properties of GS and CNTS with respect to RS are reported for each spider in Figs. 3a,b,c. Green bars indicate an enhancement of the mechanical properties, while red bars correspond to a decrease. The highest toughness modulus is ~1567 J/g, calculated assuming a mean silk density ~1.34 $g/cm^3$[8], see Fig. 4. The stress-strain curves of the silk fibres are presented in Fig. 5 and Fig. S6 in the S.I. The mean values of the Young's modulus, ultimate strain, fracture strength and toughness, are reported in Tables 1,2,3. The best mechanical performances are observed for the samples after the first collection (RDS_1). The second collection (RDS_2) does not show mechanical enhancement with respect to the first or to RS, probably due to a physiological spider weakening during segregation, since neither additional food nor water were available during the experimental period, except SWNTs and graphene dispersions. In the cases marked with an asterisk in Figs. 5b,c, the second collection was impossible since the corresponding spiders died. Note that spider 5 died after the first treated dragline silk collection, but was able to spin the silk with the maximum increment in mechanical performance, whereas spider 7 spun the silk with the highest absolute values and survived.

Fracture strength, Young's modulus and toughness increase by a factors from 3 to 6 in CNTS and between 2 and 5 for GS. The highest fracture strength and Young's modulus increments are respectively of +731% (3.9 GPa) and +1183% (37.9 GPa) for CNTS, corresponding to an increment ~+663% of toughness (1.1 GPa), with a decrement of 41% of ultimate strain (0.6 mm/mm). Note that the combination of increment in toughness and decrement in ultimate strain is very peculiar and fundamental in applications such as ballistic protection and flak jackets[20], where high performance textiles are required to stop bullets in millimeters[43]. The second highest increments were of +350% (2.0 GPa) in fracture strength and +330% (19.3 GPa) in Young's modulus for SWNT-2-CNTS, corresponding to an increment of +204% in toughness (0.4 GPa), with a decrement of 36% of ultimate strain (0.3 mm/mm). A lower, but still significant, increment was observed for GS, with +151% (1.2 GPa) in fracture strength and +142% (13.0 GPa) in Young's modulus, corresponding to an increment of both toughness and ultimate strain of +250% (0.3 GPa) and +166% (0.4 mm/mm), respectively.

These data seem to be justified only by the presence of SWNTs or graphene within the bulk of the silk fibre, combined with a very efficient natural spinning/mixing mechanism. However, we cannot exclude the presence of SWNTs and graphene on the fibre silk surface as a result of spinning in an environment containing SWNT and graphene. However, such external coating on the fibre surface is not expected to significantly contribute to the observed mechanical strengthening, which we attribute to the inclusion of SWNTs and graphene within the fibre matrix.

The larger increments in the mechanical properties for CNTS could be due to the lateral characteristic dimensions of graphene flakes (~200-300 nm), two orders of magnitude larger than the characteristic SWNT diameter, corresponding to a less efficient load transfer as resulting from (i) a longitudinal segment length of the flake not always longer than the critical length for maximal load transfer (as deduced by shear lag theory[44]), (ii) a self-crumpled configuration of the flakes, (iii) a larger misalignment with respect to the SWNT case. Finding solutions for the three previously mentioned problems should enable us to obtain GS with mechanical properties superior to SWNTs, thanks to the two surfaces available for load transfer in graphene flakes[44.] Tuning the constitutive law of the silk could also maximize the robustness of an entire structure[22] and the insertion of knots has already been demonstrated to be a powerful tool for tuning the constitutive law of high-tough fibers[8].

We perform atomistic simulations of GS and CNTS as a simple proof of concept by considering only one of the two copolymers present in the spider silk: major ampullate spidroin 2 (MASP2) (Figs. 6a,b), because it is the component delivering the major contribution to the stress-strain curve[17]. We stress that spider silks have a much more complex structure than modelled here. Nevertheless, the good agreement between computational results and experiments justifies this assumption. MAPS2 forms random coil and helical structures and has a known primary sequence[45], as discussed in the S.I. The results are reported in Figs. 5c,d and show that the inclusion of SWNTs or graphene increases toughness with respect to the pristine MASP2 protein, the effect being larger in the presence of graphene. Inspection of the numerically generated molecular dynamics trajectories shows that the SWNT and graphene surround the protein and contrast the unfolding effect of the external force $F$.

Numerical simulations of the GS and CNTS mechanical properties are also performed, as discussed in S.I. To do this, a previously developed Hierarchical Fibre Bundle Model (HFBM)[46] was employed, whereby the macroscopic fibres are modelled as networks of micro-fibres and reinforcements arranged in parallel and in series subjected to uniaxial tension, with statistically-distributed fracture strengths, according to measured or known input parameters for the constituents. The mechanical properties of the spider silk were derived from mean values in experimental measurements on RS, while those of SWNTs and graphene were obtained from literature[47,48]. Simulations are used to estimate the equivalent volume fractions ($V_f$) of reinforcements obtained in the treated silk fibres. As a reference, comparisons are made with basic rules of mixtures (RM)[48]. The calculated $E$, $\sigma$ and $\varepsilon$ values are compared to experimental results in Fig. 7. This shows that the equivalent volume fractions in GS and SWNTs vary between 1% (sample 15) and 15% (sample 7), with an average of 6.9%.

In summary, spiders placed in an environment with water solutions of nanotubes or graphene produce dragline silk with unprecedented mechanical properties, realizing the toughest achieved fibers, with a strength only comparable with that of the strongest carbon fibers or that of the limpet teeth[49]. Knots could further increase the toughness[8]. Spiders could spin graphene and nanotubes in the silk also as an efficient way of eliminating them from their organism. Spider natural and very efficient spinning can thus allow the collection of the most performing silk fiber when compared to synthetic recombinant silks, which represents the most promising silk material to be efficiently reinforced. This new reinforcing procedure could also be applied to other animals and plants, leading to a new class of bionic materials for ultimate applications.

**Acknowledgments**
We acknowledge "Nanofacility Piemonte" at the INRIM Institute for the FESEM, a Newton International Fellowship, a Royal Society Wolfson Research Merit Award, EPSRC grants EP/K01711X/1, EP/K017144/1, EP/L016087/1. N.M.P. is supported by the European Research Council (ERC StG Ideas 2011 BIHSNAM no. 279985 on 'Bio-inspired hierarchical supernanomaterials', ERC PoC 2013-1 REPLICA2 no. 619448 on 'Large-area replication of biological anti-adhesive nanosurfaces', ERC PoC 2013-2 KNOTOUGH no. 632277 on 'Super-tough knotted fibres'), by the European Commission under the Graphene Flagship (WP10 'Nanocomposites', no. 604391) and by the Provincia Autonoma di Trento ('Graphene nanocomposites', no. S116/2012-242637 and reg. delib. no. 2266).". This paper is dedicated to the memory of prof. Franco Montevecchi.

**Author contributions**
N.M.P. had the idea, designed and supervised the entire research and analyzed the results. E.L. collected the silk and perfomed the tensile tests. Fr. B. provided and characterized the graphene and carbon nanotube solutions. M.B. performed the Raman analysis, Fe. B. performed the hierarchical fibre bundle simulations. S.T. and G.G. performed the atomistic simulations. A.C.F. analyzed the Raman results. N.M.P. and E.L. drafted the manuscript and all authors contributed to the writing of its final version.

**TABLES**

**Table 1.** Mechanical properties (average values) of RS samples.

| Spider n. | Young's Modulus (GPa) | Ultimate Strain (mm/mm) | Fracture Strength (MPa) | Toughness Modulus (MPa) |
|---|---|---|---|---|
| 1 | 6.0 | 0.288 | 795.2 | 128.2 |
| 2 | 27.3 | 0.578 | 2397.2 | 713.0 |
| 3 | 13.4 | 0.464 | 1257.5 | 422.6 |
| 4 | 1.9 | 1.381 | 465.1 | 235.6 |
| 5 | 3.2 | 1.017 | 534.7 | 172.4 |
| 6 | 37.1 | 0.278 | 4045.9 | 732.1 |
| 7 | 15.1 | 0.394 | 1726.6 | 476.4 |
| 8 | 2.1 | 0.689 | 179.7 | 61.1 |
| 9 | 5.9 | 0.534 | 580.7 | 205.3 |
| 10 | 3.0 | 0.551 | 281.2 | 75.3 |
| 11 | 24.3 | 0.745 | 1969.1 | 764.7 |
| 12 | 3.1 | 0.769 | 173.4 | 48.9 |
| 13 | 5.5 | 0.879 | 648.6 | 320.3 |
| 14 | 3.8 | 1.708 | 364.0 | 247.8 |
| 15 | 9.2 | 0.256 | 825.2 | 101.8 |

**Table 2.** Mechanical properties (average values) of the first collection of silk samples produced after exposure of the spiders to the SWNT or graphene dispersions (Spiders n. 1-6 with SWNT-1, spiders n. 7-10 with SWNT-2, spiders n. 11-15 with graphene). The largest increments in the silk mechanical properties are observed for spider 5 whereas the highest absolute values are observed for spider 7.

| Spider n. | Young's Modulus (GPa) | Ultimate Strain (mm/mm) | Fracture Strength (MPa) | Toughness Modulus (MPa) |
|---|---|---|---|---|
| 1 | 3.9 | 0.622 | 326.1 | 66.7 |
| 2 | 40.1 | 0.200 | 3914.6 | 587.2 |
| 3 | 8.7 | 0.678 | 1195.8 | 387.0 |
| 4 | 2.4 | 0.946 | 579.4 | 187.6 |
| 5 | 37.9 | 0.598 | 3907.2 | 1144.0 |
| 6 | 9.6 | 0.498 | 954.3 | 210.2 |
| 7 | 47.8 | 0.749 | 5393.5 | 2143.6 |
| 8 | 3.1 | 0.406 | 315.7 | 47.8 |
| 9 | 19.3 | 0.342 | 2034.9 | 419.8 |
| 10 | 0.2 | 0.319 | 20.1 | 2.6 |
| 11 | 0.8 | 0.331 | 58.0 | 7.9 |
| 12 | 4.9 | 0.430 | 607.5 | 148.5 |
| 13 | 3.1 | 0.519 | 421.8 | 130.3 |
| 14 | 0.4 | 0.288 | 45.9 | 6.0 |
| 15 | 13.0 | 0.426 | 1245.6 | 254.7 |

**Table 3.** Mechanical properties (average values) of the second collection of silk samples produced after exposure of the spiders to the SWNT and graphene dispersions. 26% of the spiders died between the first and second collections, thus the corresponding data are absent.

| Spider n. | Young's Modulus (GPa) | Ultimate Strain (mm/mm) | Fracture Strength (MPa) | Toughness Modulus (MPa) |
|---|---|---|---|---|
| 1 | 2.8 | 0.423 | 275.1 | 55.3 |
| 2 | 9.5 | 0.368 | 906.2 | 168.1 |
| 3 | 5.1 | 0.753 | 562.5 | 157.4 |
| 4 | - | - | - | - |
| 5 | - | - | - | - |
| 6 | 3.6 | 1.325 | 507.6 | 198.1 |
| 7 | 3.6 | 0.818 | 473.1 | 195.6 |
| 8 | 9.9 | 0.509 | 1222.8 | 196.9 |
| 9 | - | - | - | - |
| 10 | 1.4 | 1.055 | 211.2 | 99.6 |
| 11 | 8.6 | 0.540 | 1102.5 | 241.4 |
| 12 | 5.6 | 1.082 | 716.7 | 286.7 |
| 13 | 5.6 | 0.577 | 499.6 | 125.7 |
| 14 | - | - | - | - |
| 15 | 2.5 | 1.138 | 216.9 | 110.4 |

# FIGURES

**Figure 1: Optical characterization of dragline silk spun by spiders after exposure to nanotube or graphene dispersions.** a) Optical picture of a spider silk fibre suspended by a mechanical support (left) and map of the 2D peak intensity overlapped to the optical picture of the spider silk spun by spiders after exposure to graphene (right). b) Raman spectra of RS (black line, at 514.5nm), GS at 514.5nm (green line) and 633nm (red line). Raman spectra of c) SWNT-1-CNTS d) SWNT-2-CNTS at 514.5nm (green line) and 633nm (red line) excitation wavelengths.

**Figure 2: Raman spectra of original graphene and nanotube samples compared to those measured in RS and SWNTs.** Raman spectra of graphene (black line) and GS (red line) at a) 514.5nm and b) 633nm. Raman spectra of c,d) SWNT-1, SWNT-2 and SWNT-1-CNTS and SWNT-2-CNTS at (c,e) 514.5nm and (d,f) 633nm excitation wavelengths.

**Figure 3: Mechanical properties of dragline silk spun by spiders after exposure to nanotube or graphene dispersions.** Percentage increment (green bars) or decrement (red bars) of the mechanical properties measured from neat dragline silk (RS) to the first collection of treated dragline silk samples with (**a**) SWNT-1-CNTS, (**b**) SWNT-2-CNTS or (**c**) GS. The symbol $^{(*)}$ specifies that spider died before the second collection.

**Figure 4: Toughness modulus and strength of different materials and composites.** Note the toughness record of the present silk and its high strength (only comparable to that of carbon fibers or limpet teeth[49]). Also, knots in this silk could further increase its toughness thanks to the concept presented in ref. 8.

**Figure 5: Stress-strain curves of the silks showing the highest mechanical properties.** (**a**) Stress-strain curves of silks that showed the highest increments of the mechanical properties after the first (RDS_1) or second (RDS_2) collection for (**b**) SWNT-1, (**c**) SWNT-2 or (**d**) GS. The symbol $^{(*)}$ specifies that spiders died before the second collection.

**Figure 6: Proof of concept via atomistic simulations of force-displacement curves.** Arrangement of (**a**) SWNTs or (**b**) graphene around MASP2 and corresponding (**c**) force-displacement curves. Dependence of applied force end-to-end distance and (**d**) corresponding dissipated energy (here called toughness) for MASP2. We report the applied force $F$ as a function of the end-to-end distance $d$, instead of the stress-strain curve, and its integral (energy) instead of the toughness modulus as we are working with a single MASP2 protein.

**Figure 7: Comparison of experiments with HFBM simulations and direct or indirect rules of mixture (RM).** RM predictions are shown as blue lines (direct RM law, taken as an upper bound for predictions) and as red lines (inverse RM law, taken as lower bound for the predictions). Experimental data are plotted as green triangles, with corresponding sample numbers, and with error bars corresponding to the experimental standard deviations (with $V_f$ values estimated from RM or HFBM, as explained below), while HFBM data are shown as hollow diamonds, with error bars corresponding to uncertainties due to statistical variations in the simulations. Only experimental data compatible with a rule of mixtures $V_f$ estimation. *i.e.*, with improved properties with respect to RS, are shown.

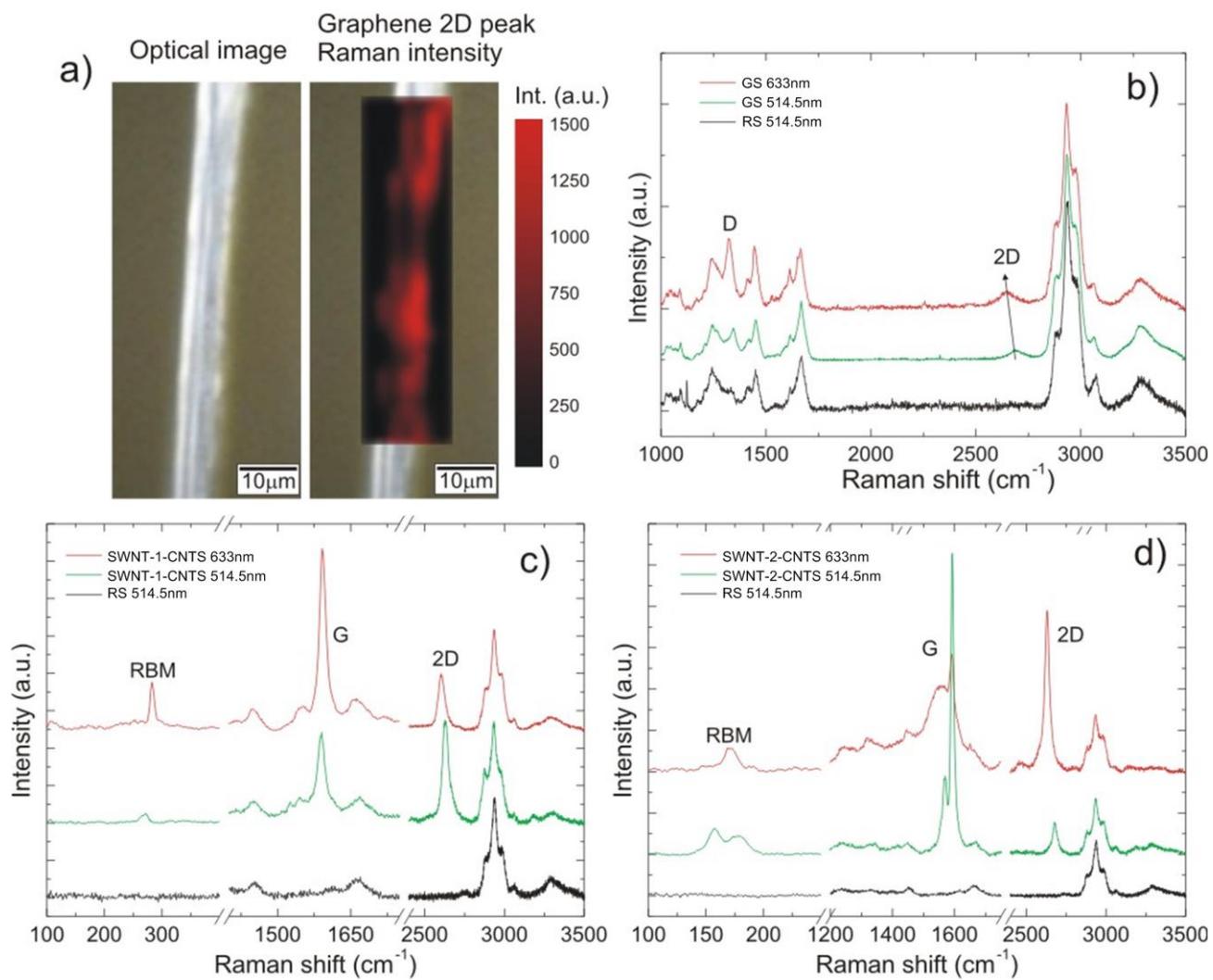

**Figure 1**

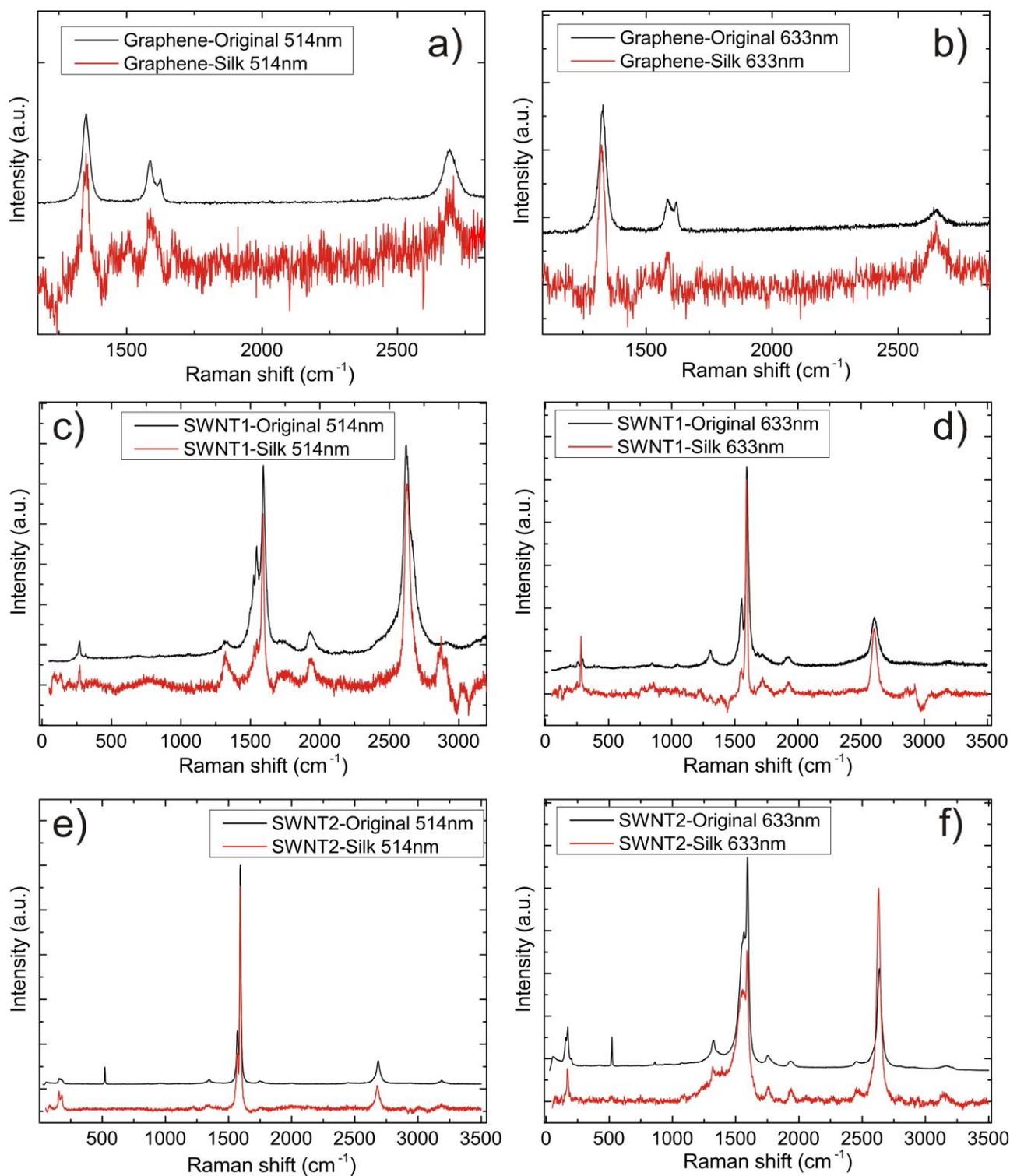

**Figure 2.**

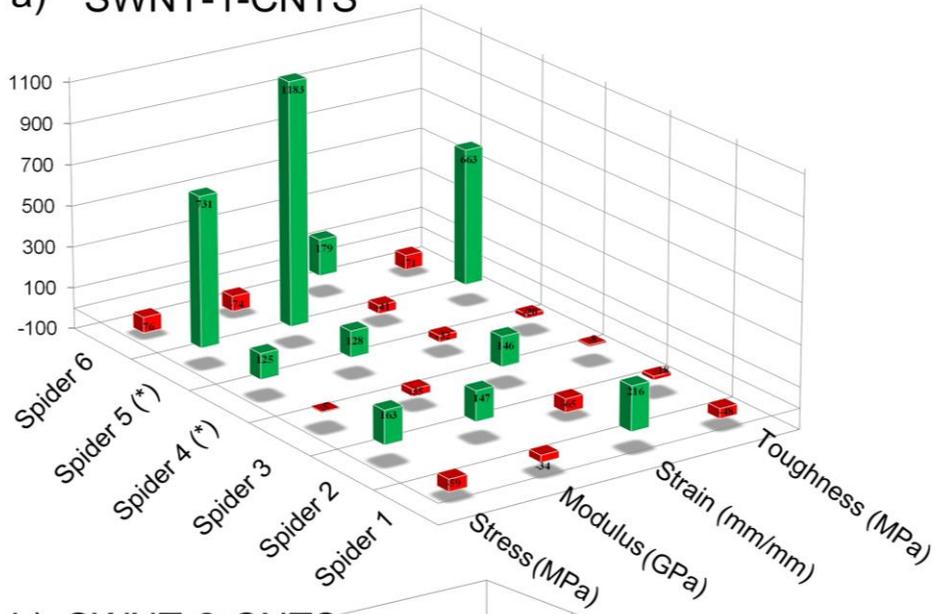
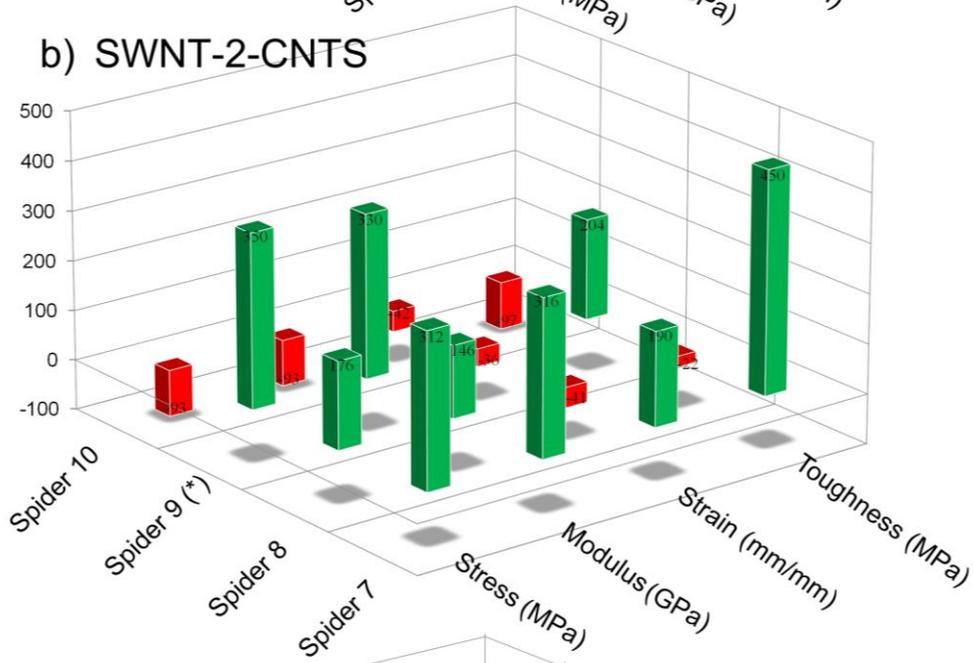
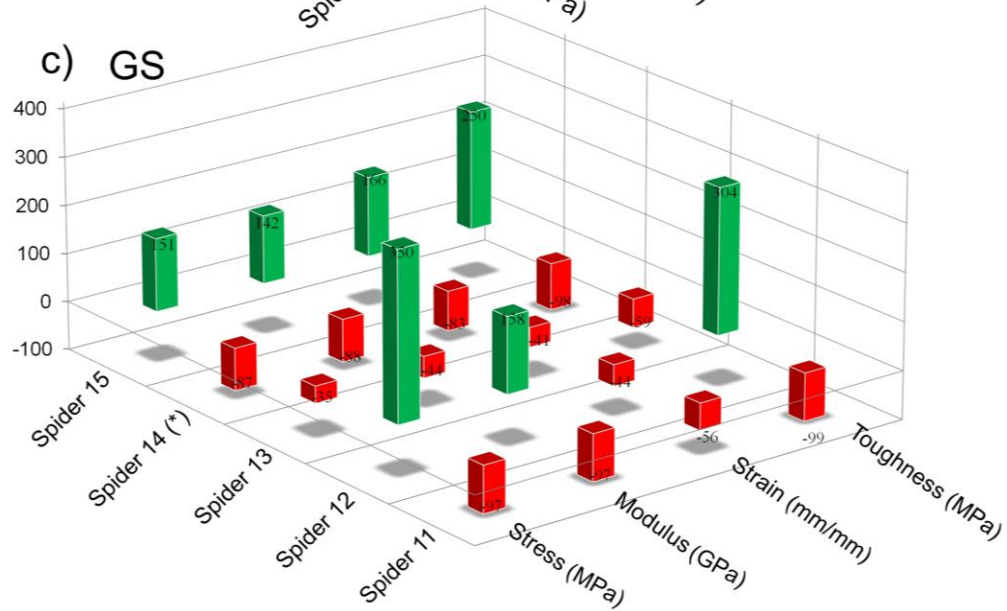

**Figure 3**

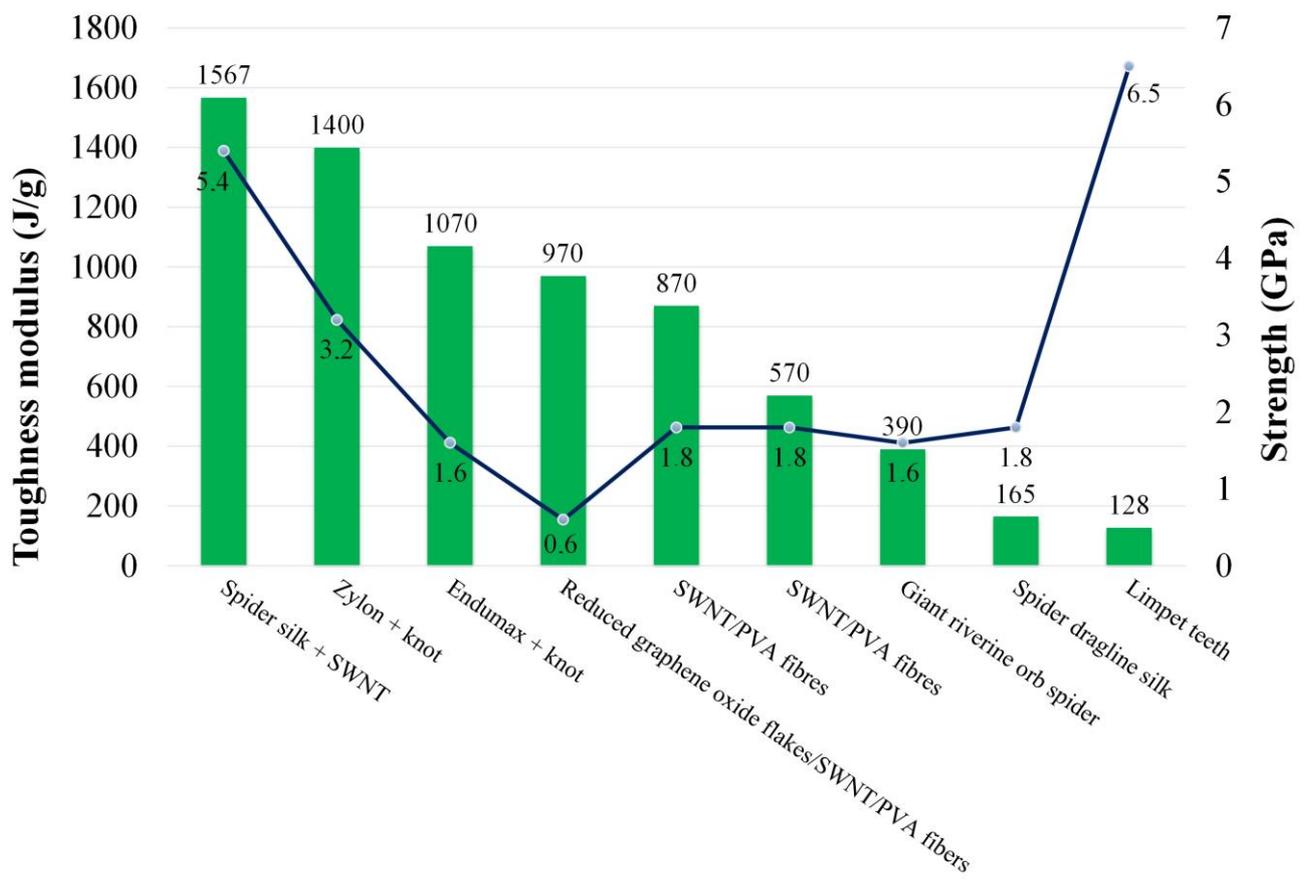

**Figure 4**

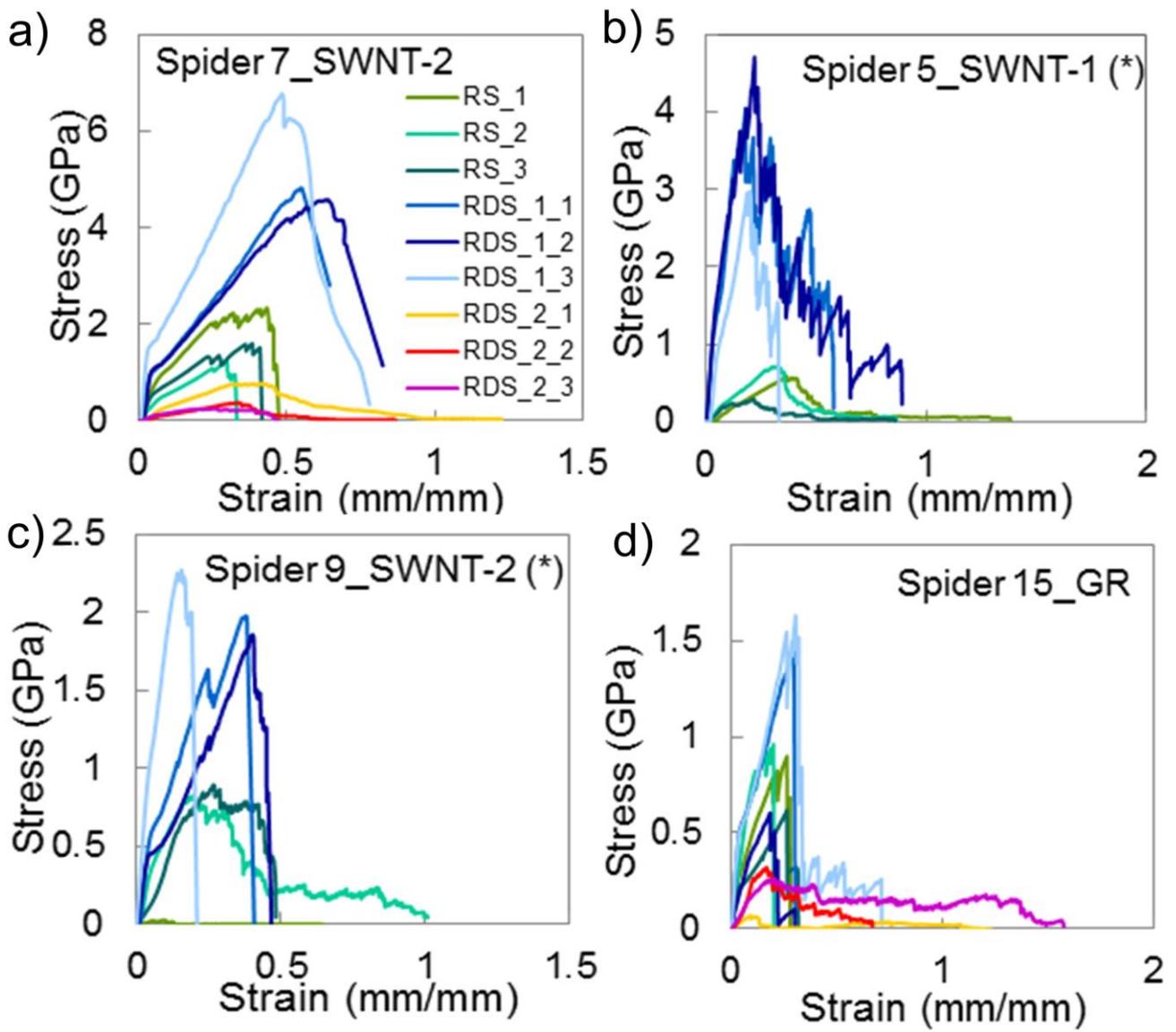

**Figure 5**

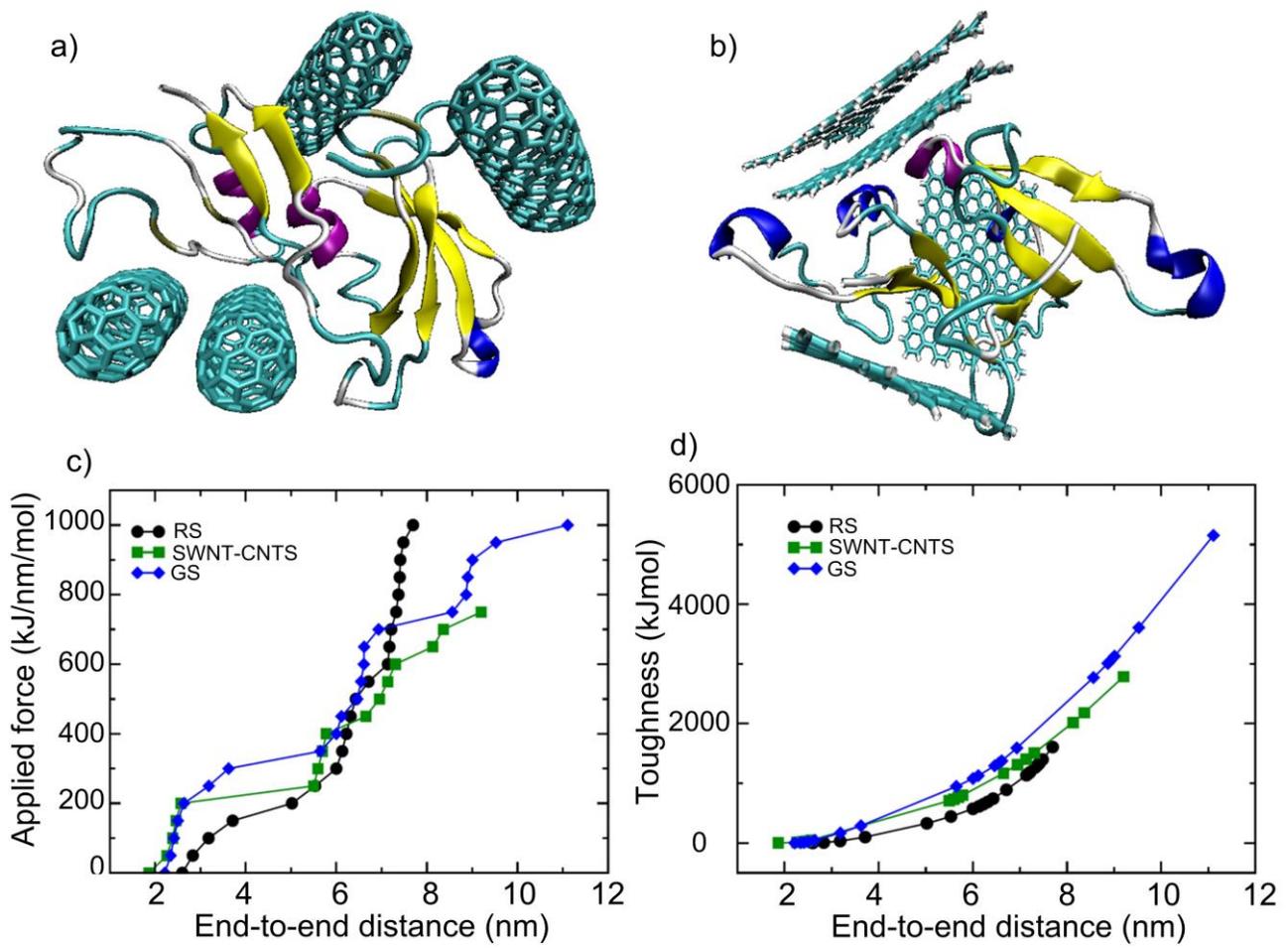

**Figure 6**

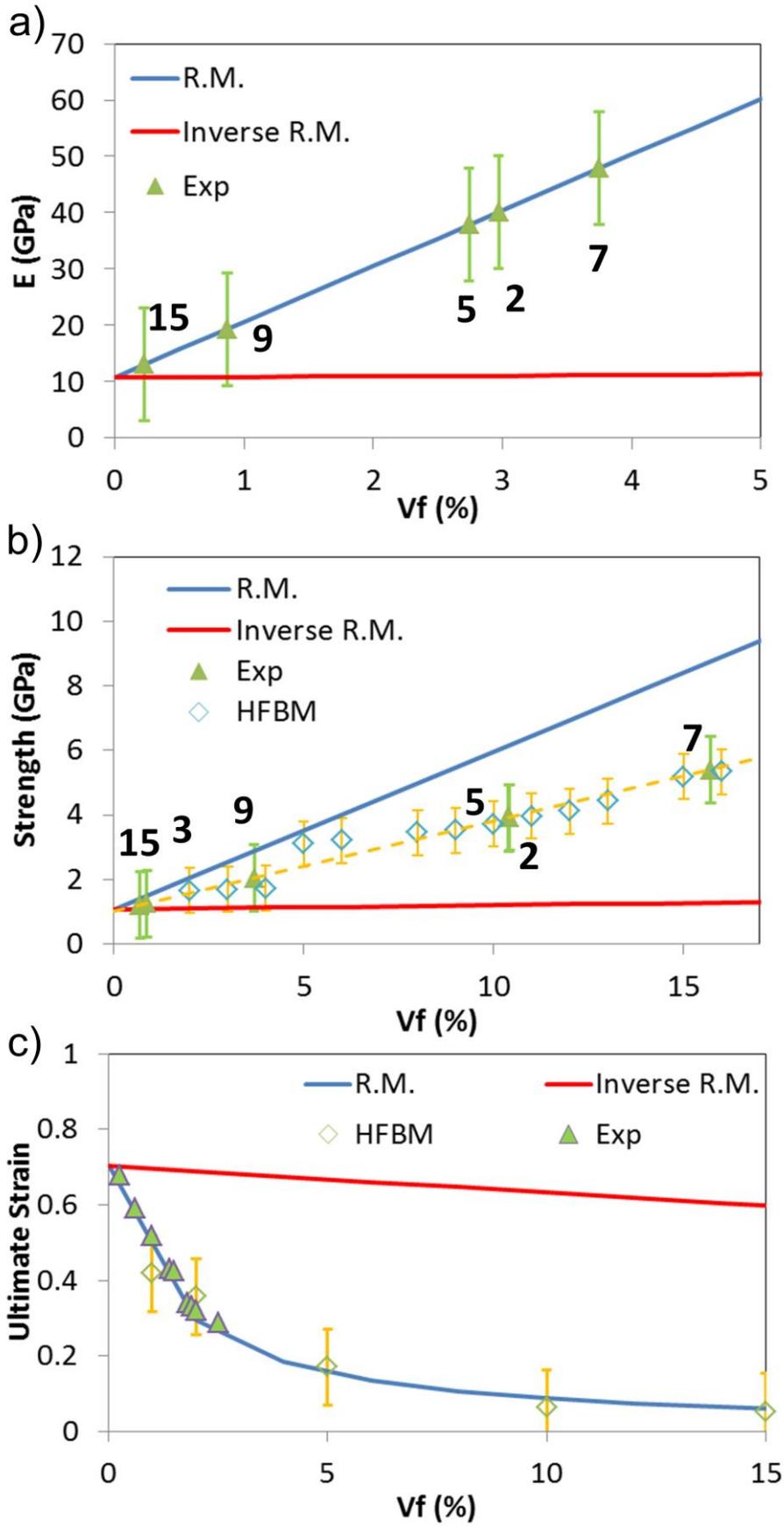

**Figure 7**

**Silk reinforced with graphene or carbon nanotubes spun by spiders**
**Supplementary Informations**

Emiliano Lepore[1], Francesco Bonaccorso[2,3], Matteo Bruna[2], Federico Bosia[4], Simone Taioli[5,6], Giovanni Garberoglio[5], Andrea. C. Ferrari[2], Nicola Maria Pugno[1,7,8*]

[1] *Laboratory of Bio-inspired & Graphene Nanomechanics, Department of Civil, Environmental and Mechanical Engineering, University of Trento, Via Mesiano 77, 38123 Trento, Italy.*
[2] *Cambridge Graphene Centre, University of Cambridge, 9 JJ Thomson Avenue, CB3 0FA, Cambridge UK.*
[3] *Istituto Italiano di Tecnologia, Graphene Labs, Via Morego 30, 16163 Genova, Italy.*
[4] *Department of Physics and "Nanostructured Interfaces and Surfaces" Centre, Università di Torino, Via P. Giuria 1, 10125, Torino, Italy.*
[5] *European Centre for Theoretical Studies in Nuclear Physics and Related Areas (ECT*), Bruno Kessler Foundation & Trento Institute for Fundamental Physics and Applications (INFN-TIFPA), Trento, Italy.*
[6] *Faculty of Mathematics and Physics, Charles University, Prague, Czech Republic.*
[7] *Centre of Materials and Microsystems, Bruno Kessler Foundation, Via Santa Croce 77, 38122 Trento, Italy.*
[8] *School of Engineering and Materials Science, Queen Mary University, Mile End Road, London E1 4NS, UK.*

**S1. Collection of spiders**

We randomly collected fifteen Pholcidae spiders (male and female of different ages). The sampling site was Torrente Chisone, between Macello and Garzigliana, Province of Torino, Italy (Geographical coordinates: 44.844 East, 7.385 North). Each spider was gently and individually segregated in a box of 19 x 12.5 x 7.5 cm$^3$ with four air inlets, a transparent top, white bottom and lateral sides. Spiders were transferred within 3 days from capture under an extractor fan with controlled ambient condition temperature and humidity (21.0 ± 0.2 °C and 54.3 ± 0.7 %) in the MicroTechnologies Laboratory (Bruno Kessler Foundation, Trento, Italy) where they were kept during the entire experimental procedure.

After 5 days reference dragline silk samples (RS) were collected. Subsequently, spiders were exposed to aqueous dispersions of graphene and single wall carbon nanotubes (SWNTs), Fig S1. Nine spiders (1, 4, 5, 7, 9, 11, 13, 14, 15) started to spin silk after 2 days, while 6 took 12 days (2, 3, 6, 8, 10, 12). The dragline silk was then collected. Spiders 1-6 were treated with SWNT-1 and 7-10 with SWNT-2 (Fig. S1c), and 11-15 with graphene (Fig. S1d). After 42 days, dragline silk was collected for the second time. The silk is in the form of a single fiber composed of multiple threads. Multiple samples are obtained by cutting the fiber into 20mm-length strands for tensile tests and for the measurement of the cross-sectional area, following the procedure described in Ref. [2]. This suggested to take adjacent samples to ensure reproducibility of fiber properties and to use one out of five samples (instead of three as here) as control to measure the fiber cross-sectional area by Field Emission Scanning Electron Microscopy (FESEM). In our work, the cross-sectional area of the fiber is determined using a FESEM (FEI-InspectTM F50, at 5-10 kV) without sputter coating.

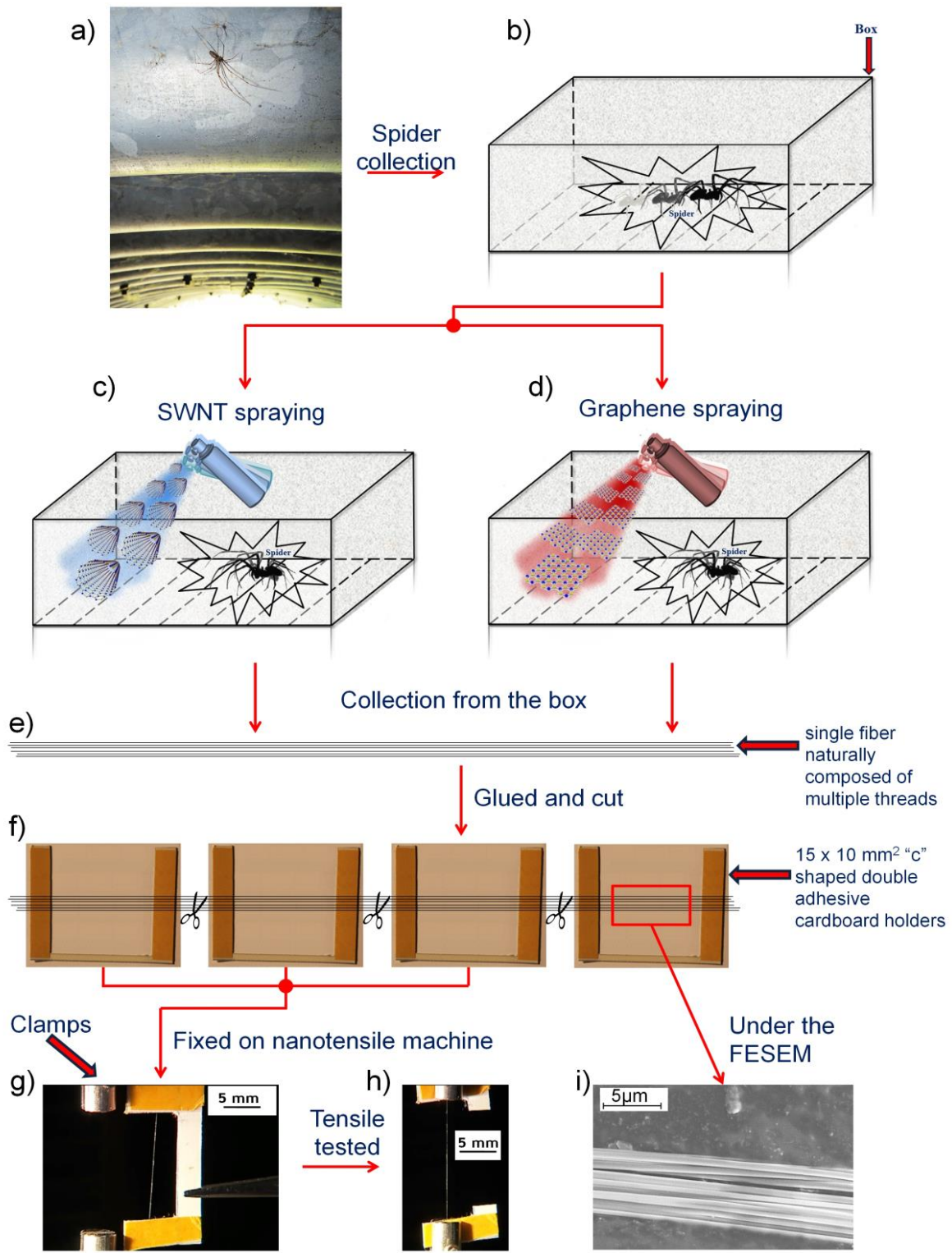

**Figure S1:** *Schematic illustration of the experimental procedure. (**a**) Sampling site and experimental box for the collection of (**b**) neat dragline silk samples and, after spraying (**c**) nanotubes or (**d**) graphene aqueous dispersions, of treated dragline silk samples. (**e**) Collection of silk in the form of a single fibre, composed of multiple threads, from which multiple samples have been obtained by fixing the ends of the fibre to 15 x 10 mm$^2$ "c" shaped double adhesive cardboard holders and (**f**) cutting the fibre into four shorter pieces, (**g**) mounted on nanotensile machine clamps (**h**) for nanotensile tests and (**i**) for measuring the fibre cross-sectional surface area with a FESEM. Scale bar 5 μm.*

## S2. Characterization of graphene and SWNT dispersions

### S2.1 Optical absorption Spectroscopy

*S2.1.1 Optical absorption spectroscopy of SWNTs*

Optical absorption spectroscopy (OAS) reveals various properties of SWNT dispersions such as transition energies,[3,4,5] bundling[3,4,6] and concentration.[7]

Optical absorption spectra were acquired in a Perkin Elmer 950 with 1nm resolution. Measurements were carried out in the range 400-1300nm, limited by the strong absorption features of water. However, this range is sufficient to cover the first and second excitonic transitions of s-SWNTs[2,8,9] and the first transition of m-SWNTs, for CoMoCAT[3,8,9] and the second and third transitions of s-SWNTs and first of m-SWNTs[10,11] for the P2 samples. Absorption from solvent and surfactants is subtracted, by measuring solutions with only solvent and surfactant.

The assignment of the optical transitions is based on the empirical Kataura plot of Ref. 21. This gives values of optical transition frequencies versus chirality for SWNTs in aqueous surfactant dispersions, and is more appropriate than Kataura plots theoretically derived from tight binding and other models[12].

The OAS of SWNT-1 and SWNT-2 are reported in Fig. S2. SWNT-1 have sharper peaks with than SWNT-2 samples.

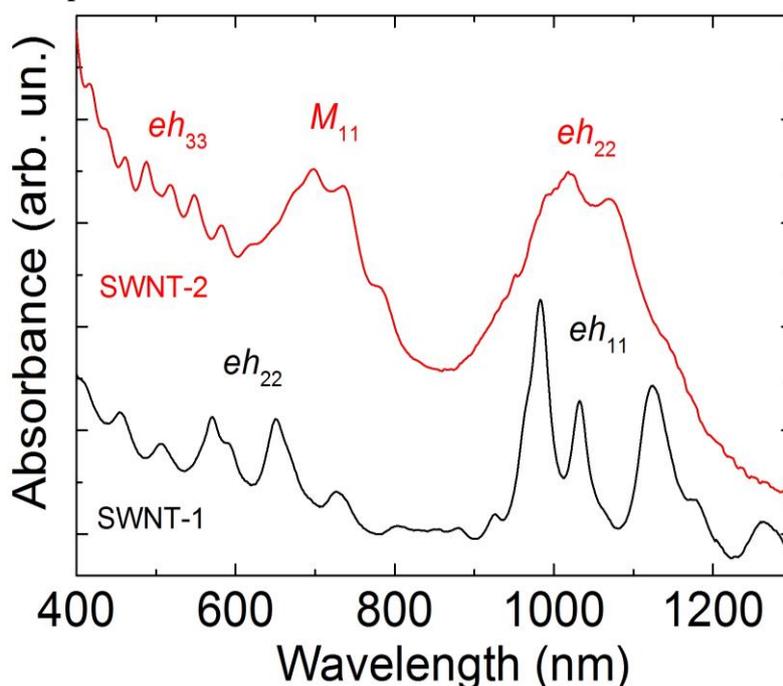

**Figure S2:** *Absorption spectra of SWNT-1 (black curve) and SWNT-2 (red curve). The labels $eh_{11}$, $eh_{22}$, $eh_{33}$ and $M_{11}$ refer to the first, second, third semiconducting and the first metallic excitonic transition, and are a guide to the eye, since overlap between different excitonic transitions exists[11,21]. The spectra are normalized for a clear visualization.*

*S2.1.1 Optical absorption spectroscopy of graphitic flakes*

Fig. S3 report the absorption spectrum of the graphene dispersion. The peak at ~266 nm is a signature of the van Hove singularity in the graphene density of states.[13] OAS used to evaluate the concentration of graphitic material in dispersion, using the experimentally derived absorption coefficient of 1390Lg$^{-1}$m$^{-1}$ at 660nm[14,15,16] we estimate concentration ~0.03mg/ml.

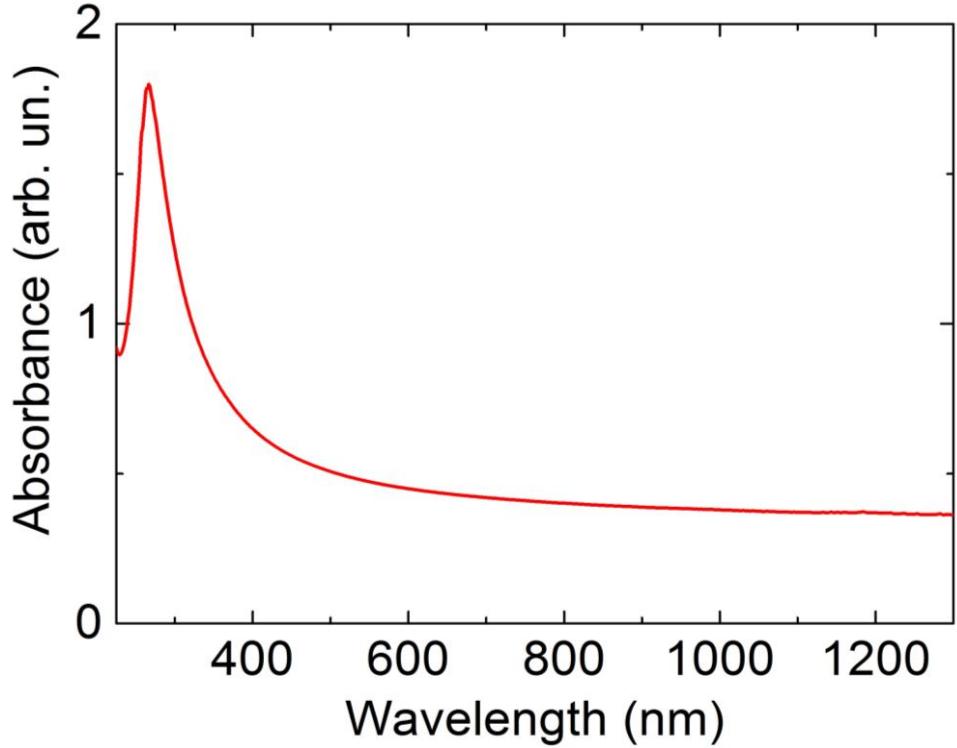

**Figure S3:** *Optical absorption spectrum of graphene dispersion in water with SDC surfactant.*

**S2.2 Raman Spectroscopy**

**S2.2.1 Pristine SWNTs**

Raman spectroscopy can be used to probe SWNTs within dispersions. In the low frequency region, the Radial Breathing Modes (RBMs) are observed[17]. Their position, Pos(*RBM*), is inversely related to SWNT diameter, $d$[18,19,20], as given by

$$Pos(RBM) = \frac{C_1}{d} + C_2.$$

Combining Pos(*RBM*), with excitation wavelength and the '*Kataura plot*'[11,21], it is, in principle, possible to derive the SWNT chirality[22,23].

Matching the diameter with excitation wavelength in the Kataura plot also gives information on the semiconducting or metallic character. A variety of $C_1$ and $C_2$ were proposed for this relation[17,18,19,23,24]. Here we use the $C_1$=214.4 cm$^{-1}$ nm and $C_2$=18.7 cm$^{-1}$, from Ref. 18. These were derived by plotting the resonance energy as a function of inverse RBM frequency without additional assumptions. We also validated our results by using the parameters proposed in Refs. 19, 24, 25.

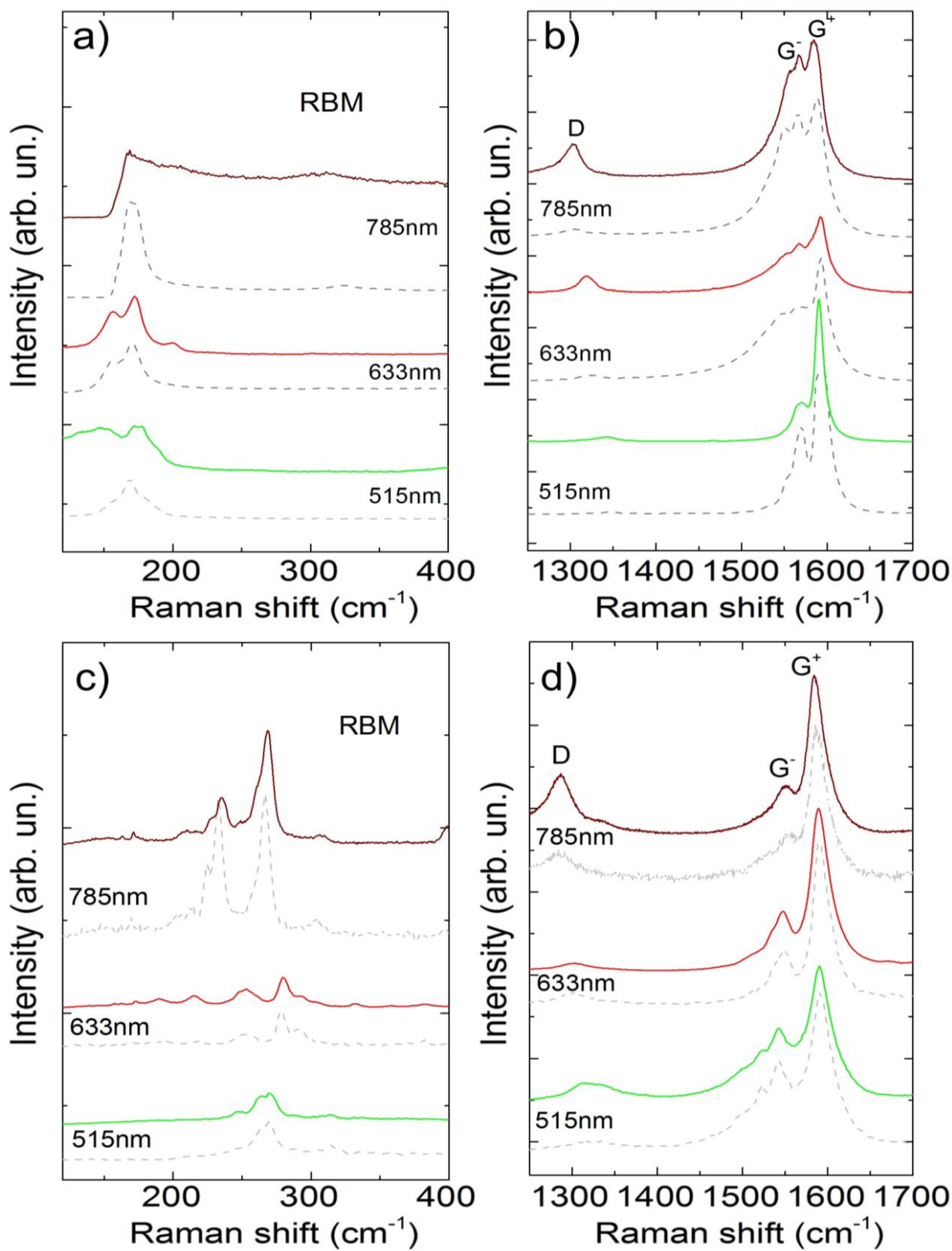

**Figure S4:** *Raman spectra of SWNTs at different excitation wavelengths: 514.5nm (green curve), 633nm (red curve) and 785nm (brown curve) (a) RBM and (b) G region for SWNT-1 and (c) RBM and (d) G region for SWNT-2. The Raman spectra of the starting materials (powders) are reported in light grey dashed lines below each curve for comparison.*

Raman spectroscopy also probes possible damage via the $D$ peak[26]. The $D$ peak is due to the breathing modes of sp$^2$ rings and requires a defect for its activation by double resonance (DR)[27,28]. The typical Raman spectrum of SWNTs in the 1500-1600 cm$^{-1}$ region consists of the $G^+$ and $G^-$ bands. In s-SWNTs, they originate from the longitudinal (LO) and tangential (TO) modes, respectively, derived from the splitting of the $E_{2g}$ phonon of graphene at the Brillouin zone centre[27,28,29,30,31]. The positions of the $G^+$ and $G^-$ peaks, Pos($G^+$), Pos($G^-$), are diameter dependent and their separation increases with decreasing diameter[32,33]. In m-SWNTs, the assignment of the $G^+$ and $G^-$ bands is the opposite, and the FWHM of the $G^-$ peak, FWHM($G^-$), is larger and Pos($G^-$) down-shifted with respect to the semiconducting counterpart[32,34]. Thus, a wide, low frequency $G^-$ is a fingerprint of m-SWNTs. On the other hand, the absence of such a feature does not necessarily imply that only s-SWNTs are present, but could signify that m-SWNTs are off-resonance.

Doping could also modify positions and FWHMs[35,36]. In m-SWNTs, a Pos($G^-$) blueshift, accompanied by a FWHM($G^-$) decrease is observed with electron or hole doping[37,38]. In s-SWNTs, doping upshifts Pos($G^+$), but does not affect FWHM($G^+$)[35,37].

Thus, a large number of excitation wavelengths are necessary for a complete characterization of SWNTs[20,24]. Nevertheless, useful information can be derived even with few excitation wavelengths.

Raman spectra are taken on both the starting materials (powder) and the dispersions, deposited by drop-casting onto an aluminium substrate to avoid any Raman background, with a Renishaw system at 514.5 nm (2.41 eV) 633 nm (1.96 eV) and 785 nm (1.58 eV), using a 100X objective and a less than 1 mW on the sample. The RBM detection is limited by the cut-off of the notch and edge filters. These are at 120, 100 and 110 cm$^{-1}$ for 514, 633 and 785nm, respectively, limiting the detection to diameters up to ~1.9nm.

The RBM spectra of the SWNT-1 in Fig. S4(a) show a distribution, spanning the 175–370 cm$^{-1}$ range both for the starting material and the dispersion, considering the peaks for the three excitation wavelengths. This RBM range corresponds to SWNTs with ~0.6–1.35nm diameter. Fig. S4(b) plots the spectra in the G region of SWNT-1. A weak $D$ band is observed [I($D$)/I($G$) = 0.13], indicating small number of defects.[28,30] These defects could be induced by the ultrasonication process because the I($D$)/I($G$) = 0.05 of the powder is lower with respect to the dispersion.

The RBM spectra of the SWNT-2 in Fig. S4(c) show a similar distribution but peaked at lower wavenumbers with respect to SWNT-1. Indeed, the RBMs span the 150–215 cm$^{-1}$ range both for the starting material and the dispersion, considering the peaks for the three excitation wavelengths. This corresponds to SWNTs with ~1.05–1.65nm diameter. Fig. S4(d) plots the spectra in the $G$ region of SWNT-1. A weak $D$ band is also observed [I($D$)/I($G$) = 0.18] for the SWNT-2 sample, indicating small number of defects.[28,30] As for the SWNT-1, also in the case of the SWNT-2 sample, these defects could be induced by the ultrasonication process because the I($D$)/I($G$) = 0.08 of the powder is lower with respect to the one of the dispersion.

**S2.3 Pristine graphene flakes and dispersions**

The ultracentrifuged dispersions are drop-cast onto a Si wafer with 300nm thermally grown SiO$_2$ (LDB Technologies Ltd.). These samples are then used for Raman measurements at 488, 514.5, and 633nm. The $G$ peak dispersion is defined as Disp($G$) =ΔPos($G$)/Δλ$_L$, where λ$_L$ is the laser excitation wavelength.

Fig. S5a plots a typical Raman spectrum of the flakes prepared on Si/SiO$_2$ substrates.

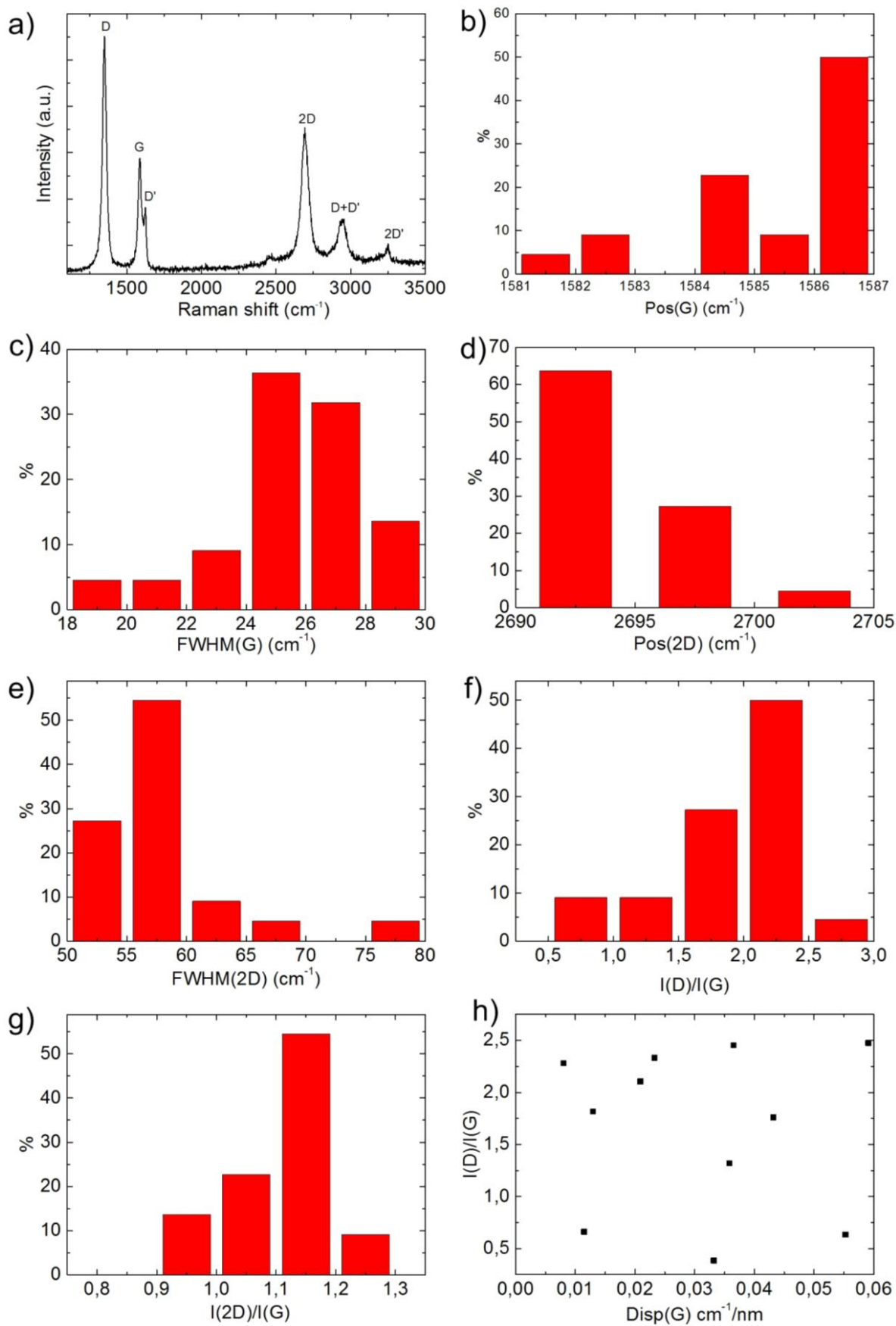

**Figure S5:** *a) Raman spectrum measured at 514.5nm excitation for a representative flake obtained via LPE of graphite. Distribution of b) Pos(2D), c) FWHM(2D), d) Pos(G), e) FWHM(G), f) I(D)/I(G), g) I(2D)/I(G), and h) distribution of I(D)/I(G) as a function of Disp(G).*

Besides the *G* and *2D* peaks, this spectrum shows significant *D* and *D'* intensities and the combination mode *D+D'*. Statistical analysis of the spectra shows that Pos(*G*) (Fig. S5b) and FWHM(*G*) (Fig. S5c) are ~1582 and ~27cm$^{-1}$. Pos(*2D*) peaks ~2695cm$^{-1}$ (Fig. S5d), while FWHM(*2D*) varies from 50 to 80cm$^{-1}$ (Fig. S5e) with a peak at 57cm$^{-1}$. This is consistent with the samples being a combination of single- (SLG) and few-layer (FLG) graphene flakes. The Raman spectra show significant *D* and *D'* intensities, with the intensity ratio I(*D*)/I(*G*) having a maximum at 2.25 (Fig. S5f). This high I(*D*)/I(*G*) is attributed to the edges of our sub-micrometer flakes[39], rather than to the presence of a large amount of structural defects within the flakes. This observation is supported by the low Disp(*G*)<0.06 cm$^{-1}$/nm, much lower than what expected for disordered carbon[28]. Combining I(*D*)/I(*G*) with Disp(*G*) allows us to discriminate between disorder localized at the edges and disorder in the bulk. In the latter case, a higher I(*D*)/I(*G*) would correspond to higher Disp(*G*). Fig. S5h show that Disp(*G*) and I(*D*)/I(*G*) are not correlated, an indication that the major contribution to the *D* peak comes from the edges of the sample.

**S.3 Nanotensile Tests**

Nanotensile tests were performed under controlled laboratory conditions (in the Laboratory of Bio-Inspired and Graphene Nanomechanics of the University of Trento), since any change in temperature and humidity may affect the mechanical characteristics of the silk threads[2,40,41,42,43].

We monitored the experimental ambient conditions with a datalogger (EL-USB-2, Lascar Electronics): the air temperature and the relative humidity were recorded to be 22.8 ± 1.3 °C and 59.1 ± 5.0 % during tensile tests. We fixed silk fibre ends to 15 x 10 mm$^2$ "c" shaped cardboard holders, with double adhesive faces. These allowed the fibre to be suspended and mounted on the nanotensile testing machine, while maintaining its original tension without damage (Fig. S6g). The tests were conducted using a nanotensile testing system (T150, Agilent, Santa Clara, USA), equipped with 500 mN maximum cell load. This can generate load-extension data with a load resolution of 50 nN and a 0.1nm displacement resolution. The cardboard holders were placed between the clamps. Once the holders were in place, the clamps were closed and one side of the holders was cut (Fig. S6h), leaving the fibre free between the clamps. We performed a continuous dynamic analysis of the silk by imposing an oscillating dynamic strain up to failure of the thread. We used a dynamic strain oscillation with a 20Hz frequency and a 0.1mN dynamic force amplitude, which enabled mechanical properties to be determined continuously.

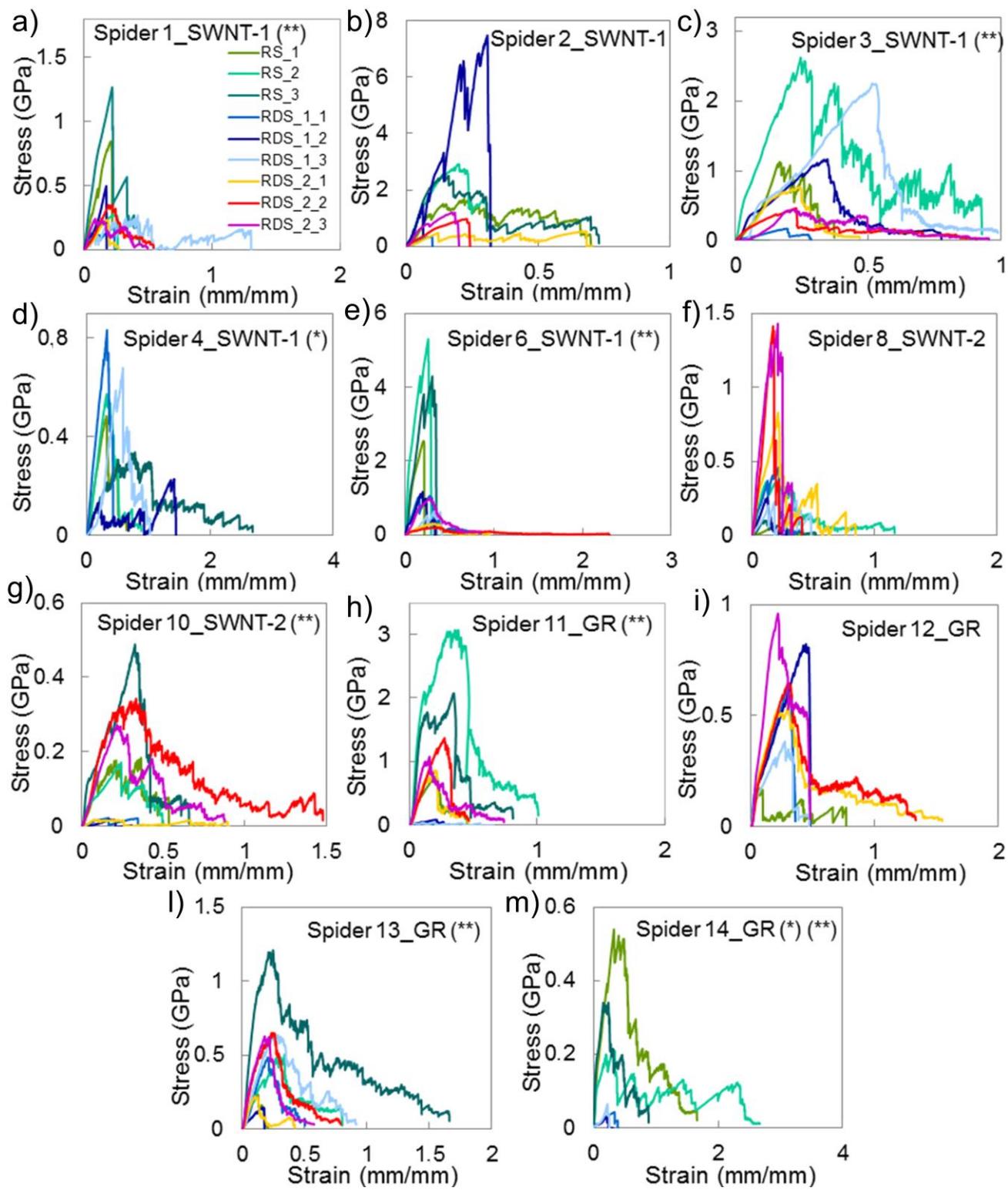

**Figure S6:** *Stress-strain curves.* Stress-strain curves for silk produced by spiders after first (RDS_1) or second (RDS_2) ingestion of a-g) SWNT-1, SWNT-2 or h-m) graphene dispersions. The symbol "**" specifies enhanced mechanical properties (fracture strength and Young's modulus) after first collection. The symbol "*" specifies that spiders died before the second collection. The legend is reported for the Spider 1_SWNT-1.

## S.4 Atomistic simulations

The three-dimensional structure of the crystalline region of major ampullate spidroin 2 (MASP2) is generated using the (PS)2-v2 program[44]. Simulations are performed using the GROMACS package (version 4.6.3)[45]. The MASP2 protein is simulated using the all-atom AMBER/99 force field.[46] We generate capped (10,10) nanotubes with a length of 2.1 nm, and squared graphene sheets (with dangling bonds saturated with hydrogen) of size 1.8 x 2.5 nm$^2$. Both nanotubes and graphene are simulated using the DREIDING force field[47] for the bonded interactions, which has been proved to be a good compromise between simplicity and accuracy for carbon allotropes, while Steele parameters are used for the Lennard-Jones carbon interactions as they are more accurate than the original DREIDING ones[48]. The starting configurations are generated by placing 4 nanotubes or 4 graphene sheets around the protein, and pulling them towards MASP2's center of mass. After removing the pulling force, we equilibrated the system for 500 ps, with all the bonds treated as flexible and using a time step of 1 fs as common practice in molecular mechanics simulations of molecules. The resulting configurations are shown in Figs. 6(a,b). We then apply a fixed force $F$ on the two protein terminals and directed along the vector connecting their centers of mass. The computations are run for 500 ps, of which 250 ps are necessary to reach equilibration, and the remaining 250 ps were used to compute average values, such as the end-to-end distance of the MASP2 protein. The force $F$ is varied in the range 50 to 1000 kJ/nm/mol in steps of 50 kJ/nm/mol. The highest value of the force was set in such a way to obtain completely stretched MASP2 configurations.

## S.5 Statistical data analysis and numerical HFBM simulations

Analysis of experimental data discussed in the Main Text and in the "Nanotensile Tests" Section indicates that there is a large statistical spread in the measured mechanical properties of RS as well as CNTS and GS. Therefore, to evaluate the effect of SWNT and graphene addition it is necessary to perform a statistical analysis of the experimental data. To do this, we adopt the Weibull distribution, which is widely used in fracture mechanics and particularly suitable to describe dispersion of mechanical properties[49], and separately fit the data relative to RS, SWNT-1-CNTS, SWNT-2-CNTS and RS. The resulting 2-parameter Weibull distributions are shown in Fig. S7. The parameters for the corresponding Weibull distributions of RS, CNTS and GS are summarized in Table S1. Scale parameters are indicative of average values, whilst shape parameters are correlated to the dispersion of the distributions. The addition of SWNTs leads to an average increase in the stiffness and strength of the silk, and an average reduction in the ultimate strain, whilst the addition of graphene does not improve average stiffness and strength, although it leads to the same decrease in average ultimate strain. In all cases, we get an increase in the dispersion (*i.e.*, decrease in Weibull shape parameters) of the mechanical properties, possibly due to variable reinforcement volume fractions. Numerical simulations of deformation and fracture are performed using a previously developed Hierarchical Fibre Bundle Model (HFBM)[50], whereby the macroscopic fibres are modelled as networks of microfibres and/or reinforcements arranged in parallel and in series, subjected to uniaxial tension. The microfibers are treated as elastic springs with statistically distributed fracture strengths, according to measured or known input parameters for the constituents (i.e. the experimentally-determined Weibull parameters from Table S.1 for the silk, and known mechanical properties from the literature for SWNT and graphene). Simulations are carried out in a multiscale procedure, from single SWNT (diameter ~1.5 nm, length ~250 nm) scale to the macroscopic scale of specimens used in experimental tests (diameter ~ 5 μm, length ~1 cm). We consider typical values for SWNT and graphene properties, respectively: Young's modulus $E_{CNT} = E_G = 1$ TPa[51,52] and strength $\sigma_{CNT} = 45$ GPa[50], $\sigma_G = 130$ GPa[51]. The experimentally obtained RS Weibull distributions are used as input properties. Fig. S8 shows typical simulations results for stress-strain curves with different SWNT volume fractions $V_f$, assuming a uniform SWNT dispersion at the lowest hierarchical simulation level (i.e. that at which individual SWNTs coincide with springs in the bundle). The figure shows that there is an overall strength and modulus increase with increasing $V_f$, while there is a corresponding ultimate strain reduction.

**Table S1.** Mean Young's modulus, strength and ultimate strain of RS, CNTS and GS, and corresponding Weibull parameters.

| Young's Modulus | Mean (GPa) | STD (GPa) | Shape parameter | Scale parameter (GPa) |
|---|---|---|---|---|
| RS | 10.73 | 10.43 | 1.19 | 11.06 |
| SWNT-1-CNTS | 17.10 | 15.70 | 0.99 | 17.70 |
| SWNT-2-CNTS | 17.60 | 18.89 | 0.48 | 14.23 |
| GS | 4.44 | 4.58 | 0.83 | 4.32 |

| Strength | Mean (MPa) | STD (MPa) | Shape parameter | Scale parameter (MPa) |
|---|---|---|---|---|
| RS | 1082.94 | 1082.94 | 1.27 | 1126.55 |
| SWNT-1-CNTS | 1401.36 | 1508.48 | 1.13 | 1953.83 |
| SWNT-2-CNTS | 834.25 | 2136.60 | 0.47 | 1523.46 |
| GS | 3299.16 | 440.94 | 0.76 | 486.08 |

| Ultimate Strain | Mean (mm/mm) | STD (mm/mm) | Shape parameter | Scale parameter (mm/mm) |
|---|---|---|---|---|
| RS | 0.70 | 0.40 | 2.09 | 0.79 |
| SWNT-1-CNTS | 0.59 | 0.22 | 2.16 | 0.69 |
| SWNT-2-CNTS | 0.45 | 0.17 | 2.57 | 0.52 |
| GS | 0.40 | 0.08 | 4.95 | 0.43 |

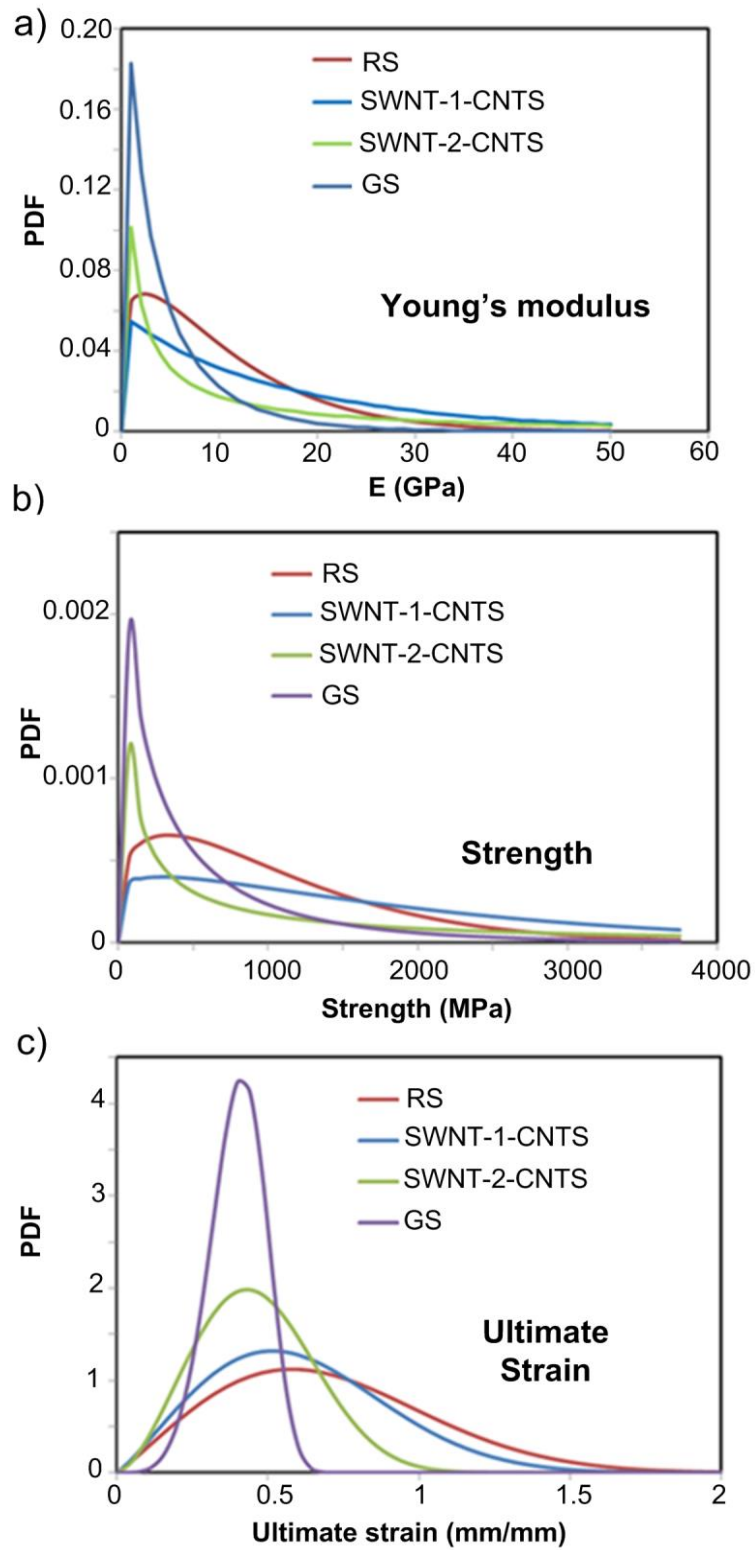

**Figure S7: Weibull fits.** Weibull fits (Probability Density Functions, PDF) on experimental data for a) Young's modulus, b) strength and c) ultimate strain of RS, CNTS, and GS.

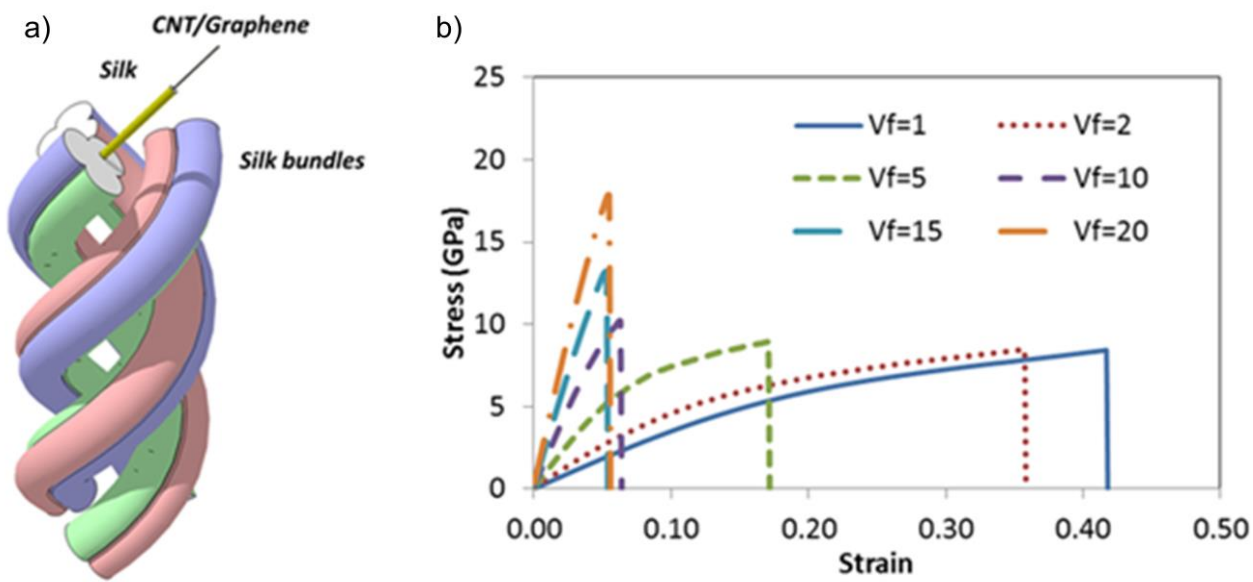

**Figure S8: HFBM simulations.** a) Schematic of the model and b) examples of stress-strain results for different volume ratios.

**References SI**